# Double PN Benchmark Solution for the 1D Monoenergetic Neutron Transport Equation in Plane Geometry

**Barry Ganapol[1]**

[1] *Department of Aerospace and Mechanical Engineering, University of Arizona;* Ganapol@cowboy.ame.arizona.edu

**Abstract:** As more and more numerical and analytical solutions to the linear neutron transport equation become available, verification of numerical results is increasingly important. This presentation concerns the development of another benchmark for the linear neutron transport equation. There are numerous ways of solving the transport equation, such as the Wiener-Hopf method based on analyticity, method of singular eigenfunctions, Laplace and Fourier transforms and analytical discrete ordinates, which is arguably one of the most straightforward, to name a few. Another potential method is the PN method, where the solution is expanded in terms of full range orthogonal Legendre polynomials and with orthogonality and truncation, the moments form a set of second order ODEs. Because of the half-range boundary conditions for incoming particles however, full range Legendre expansions are inaccurate near material discontinuities. For this reason, a double PN (DPN) expansion is more appropriate, where the incoming and exiting flux distributions are expanded separately to preserve the discontinuity at material interfaces. Here, a new method of solution for the DPN equations is proposed and demonstrated for an isotropically scattering medium.

## 1. Introduction

Boltzmann's equation of particle transport, indeed presents a significant challenge and noteworthy opportunity to solve because of its complexity and wide range of physical phenomena it describes. Originally, the non-linear integro-differential equation, as prescribed by kinetic theory of particle motion, was considered unsolvable. With time however, and advances in mathematics and physical applications, where, in some cases, nonlinearity could be relaxed to give a linear equation, the situation changed. In the early to mid twentieth century, a flurry of analytical solutions were constructed for the linear and linearized Boltzmann equation primarily based on solving partial differential equations (PDEs) with distributions admitted, specifically for one-dimension. Alongside the development of analytical solutions were numerical solutions as well such as Monte Carlo, thus enabling the practical use of Boltzmann's equation in nuclear physics experiments and nuclear reactor physics. With numerical solutions and applications to sensitive physics, there arose a need for numerical method's verification, which led to the development of benchmarks and benchmarking. This, in turn, led to a host of numerical benchmark solutions to more relevant transport applications requiring sophisticated benchmarking techniques but still generally limited to model problems. Some readers may have the misconception that benchmarking comprehensive transport algorithms is a wasted exercise since only idealized cases can be treated; however, the opposite is true. Even the most advanced numerical method to solve the transport equation can contain unknown errors that a benchmark, even for a simple problem, can easily find. The following is about developing another high precision benchmark in one-dimension.

## 2. Proposed DPN algorithm

The PN solution is a well-known solution to determine the angular flux as described by the transport equation. In the method, one expresses the angular flux to the transport equation in a 1D-plane slab as the full-range PN infinite series approximation. However, this approach is problematic for two fundamental reasons. The first concerns how the equations are closed since there is always one more unknown than equation. The second is how one best represents half-range boundary conditions by a continuous full-range expansion, which is not possible in an exact way. There have been several methodologies proposed on how to treat these outstanding issues [See Refs. 1-4]. The simplest approach to specifying boundary conditions is to avoid the difficulty altogether by resorting to the DPN moments approximation, where the expansions are split between forward and backward neutron directions. Now, the moments over the incoming fluxes on the slab near and far boundaries are exact in the limit. Since the DPN approximation also has the advantage of being a coupled set of ODEs for flux moments, one can solve for them following well-established methods. Furthermore, one way to close the system is through convergence acceleration as *N* approaches infinity.

### 2.1. DPN moment and parity equations

We consider the simplest case of 1D plane geometry where scattering is isotropic, for which the transport equation is

$$\left[\mu\frac{\partial}{\partial x}+1\right]\phi(x,\mu)=\frac{\omega}{2}\int_{-1}^{1}d\mu'\phi(x,\mu')+Q(x,\mu) \tag{1a}$$

$$\phi(0,\mu)=f(\mu)$$
$$\phi(a,-\mu)=0. \tag{1b}$$

$\phi(x,\mu)$ is the neutron angular flux, $x$ is its position, $\mu$ is the neutron direction cosine, $\omega$ is the probability of scattering and $Q$ is an imposed external source. The DPN approximation assumes the flux representation

$$\phi(x,\mu)=\begin{cases}\sum_{l=0}^{\infty}(2l+1)\phi_l^{+}(x)P_l(2\mu-1), & \mu\geq 0\\ \sum_{l=0}^{\infty}(2l+1)\phi_l^{-}(x)P_l(2\mu+1), & \mu\leq 0,\end{cases} \tag{2a,b}$$

where $P_l(2\mu\pm1)$ is the half-range Legendre polynomial of degree *l*. When projected over half-range polynomials $P_j(2\mu\pm1)$ over intervals $[0,1]$ and $[-1,0]$ respectively and from orthogonality

$$\int_0^1 d\mu P_j(2\mu\pm1)P_l(2\mu\pm1)=\frac{1}{2l+1}\delta_{j,l}, \tag{3}$$

the coefficients in Eqs(2) become the moments over the positive and negative directions,

$$\phi_j^{\pm}(x)=\int_{(0,-1)}^{(1,0)} d\mu P_j(2\mu\mp1)\phi(x,\mu). \tag{4$^\pm$}$$

The integral on the RHS of Eq(1a), called the scalar flux, in terms of moments is

$$\phi(x)=\int_{-1}^{1}d\mu'\phi(x,\mu')=\int_{0}^{1}d\mu'\phi(x,\mu')+\int_{-1}^{0}d\mu'\phi(x,\mu')=\phi_0^{+}(x)+\phi_0^{-}(x) \tag{5}$$

and the transport equation to solve becomes

$$\left[\mu\frac{\partial}{\partial x}+1\right]\phi(x,\mu)=\frac{\omega}{2}\left[\phi_0^+(x)+\phi_0^-(x)\right]+Q(x,\mu). \tag{6}$$

To find a recurrence relation for the moments, we first multiply Eq(6) over $P_j(2\mu\mp1)$

$$\begin{aligned}&\left[\frac{\partial}{\partial x}\mu P_j(2\mu\mp1)+P_j(2\mu\mp1)\right]\phi(x,\mu)=\\&=\frac{\omega}{2}\left[\phi_0^+(x)+\phi_0^-(x)\right]P_j(2\mu\mp1)+Q(x,\mu)P_j(2\mu\mp1);\end{aligned} \tag{7a$^\pm$}$$

and from the recurrence of half-range Legendre polynomials, substitute

$$\mu P_j(2\mu\mp1)=\frac{1}{2}\left[\begin{aligned}&\frac{j+1}{2j+1}P_{j+1}(2\mu\mp1)+\\&+\frac{j}{2j+1}P_{j-1}(2\mu\mp1)\pm P_j(2\mu\mp1)\end{aligned}\right], \tag{7b$^\pm$}$$

to find

$$\begin{aligned}&\frac{1}{2}\frac{\partial}{\partial x}\left[\frac{j+1}{2j+1}P_{j+1}(2\mu\mp1)+\frac{j}{2j+1}P_{j-1}(2\mu\mp1)\pm P_j(2\mu\mp1)\right]\phi(x,\mu)+\\&+P_j(2\mu\mp1)\phi(x,\mu)=\frac{\omega}{2}\left[\phi_0^+(x)+\phi_0^-(x)\right]P_j(2\mu\mp1)+Q(x,\mu)P_j(2\mu\mp1).\end{aligned} \tag{7c$^\pm$}$$

Finally, on projection over intervals $[0,1]$ and $[-1,0]$ respectively applying orthogonality

$$\begin{aligned}&\frac{1}{2}\left[\frac{j+1}{2j+1}\frac{d\phi_{j+1}^\pm(x)}{dx}+\frac{j}{2j+1}\frac{d\phi_{j-1}^\pm(x)}{dx}\pm\frac{d\phi_j^\pm(x)}{dx}\right]+\phi_j^\pm(x)=\\&=\frac{\omega}{2}\left[\phi_0^+(x)+\phi_0^-(x)\right]\delta_{j,0}+Q_j^\pm(x).\end{aligned} \tag{7d$^\pm$}$$

To arrive at the final result, change *j* to *l* and multiply by $(\pm1)^l$ to give for $l=0,1,\ldots,N-1$

$$\begin{aligned}&\frac{1}{2}\left[\frac{l+1}{2l+1}\frac{d\tilde{\phi}_{l+1}^\pm(x)}{dx}+\frac{l}{2l+1}\frac{d\tilde{\phi}_{l-1}^\pm(x)}{dx}+\frac{d\tilde{\phi}_l^\pm(x)}{dx}\right]\pm\phi_l^\pm(x)=\\&=\pm\frac{\omega}{2}\left[\tilde{\phi}_0^+(x)+\tilde{\phi}_0^-(x)\right]\delta_{l,0}\pm\tilde{Q}_l^\pm(x).\end{aligned} \tag{7e$^\pm$}$$

when

$$\tilde{\phi}_l^\pm(x)\equiv(\pm1)^l\phi_l^\pm(x), \tag{7f$^\pm$}$$

and for closure, we assume

$$\frac{d\tilde{\phi}_{N+1}^{\pm}(x)}{dx} \equiv 0. \tag{7g}^{\pm}$$

By defining the vectors

$$\begin{aligned} \tilde{\boldsymbol{\phi}}^{\pm}(x) &\equiv \left[\tilde{\phi}_0^{\pm} \quad \tilde{\phi}_1^{\pm} \quad \ldots \quad \tilde{\phi}_N^{\pm}\right]^T \\ \tilde{\boldsymbol{Q}}^{\pm}(x) &\equiv \left[\tilde{\boldsymbol{Q}}_0^{\pm} \quad \tilde{\boldsymbol{Q}}_1^{\pm} \quad \ldots \quad \tilde{\boldsymbol{Q}}_N^{\pm}\right]^T, \end{aligned} \tag{8a,b}^{\pm}$$

Eq(7e) becomes

$$\frac{1}{2}\boldsymbol{A}\frac{d\tilde{\boldsymbol{\phi}}^{\pm}(x)}{dx} \pm \tilde{\boldsymbol{\phi}}^{\pm}(x) = \pm\frac{\omega}{2}\left[\tilde{\phi}_0^{+}(x) + \tilde{\phi}_0^{-}(x)\right]\boldsymbol{1}_0 \pm \tilde{\boldsymbol{Q}}^{\pm}(x), \tag{9a}^{\pm}$$

where

$$\begin{aligned} \boldsymbol{A} &\equiv \left\{\delta_{l-1,j} l/(2l+1),\ \delta_{l,j},\ \delta_{l+1,j}(l+1)/(2l+1);\ l,j = 0,..,N-1\right\} \\ \boldsymbol{1}_0 &\equiv \{1\ 0\ 0....0\}^T. \end{aligned} \tag{9b,c}$$

Expressing the RHS more conveniently in terms of $\boldsymbol{\psi}(x)$ gives

$$\frac{1}{2}\boldsymbol{A}\frac{d\tilde{\boldsymbol{\phi}}^{\pm}(x)}{dx} \pm \tilde{\boldsymbol{\phi}}^{\pm}(x) = \pm\frac{\omega}{2}\boldsymbol{\delta}\left[\tilde{\boldsymbol{\phi}}^{+}(x) + \tilde{\boldsymbol{\phi}}^{-}(x)\right] \pm \tilde{\boldsymbol{Q}}^{\pm}(x), \tag{10a}^{\pm}$$

where the matrix $\boldsymbol{\delta}$ is

$$\boldsymbol{\delta} \equiv \boldsymbol{1}_0\boldsymbol{1}_0^T = \left\{\delta_{0,0}\right\}. \tag{10b}$$

The key to solving for the flux vector are the even-odd parity equations derived next.

### 2.2. The even/odd DPN parity equations

We form the even and odd parity moment vectors

$$\begin{aligned} \boldsymbol{\psi}(x) &\equiv \tilde{\boldsymbol{\phi}}^{+}(x) + \tilde{\boldsymbol{\phi}}^{-}(x) \\ \boldsymbol{\chi}(x) &\equiv \tilde{\boldsymbol{\phi}}^{+}(x) - \tilde{\boldsymbol{\phi}}^{-}(x) \end{aligned} \tag{11a,b}$$

with the flux moments recovered from

$$\tilde{\boldsymbol{\phi}}^{\pm}(x) = \frac{1}{2}\left[\boldsymbol{\psi}(x) \pm \boldsymbol{\chi}(x)\right]. \tag{11c}^{\pm}$$

Similarly, the source parity moments are

$$\boldsymbol{q}^{\pm}(x) \equiv \boldsymbol{Q}^{+}(x) \pm \boldsymbol{Q}^{-}(x). \tag{11d}^{\pm}$$

The parity equations come from adding and subtracting Eqs(10a)± to give

$$\frac{1}{2}\boldsymbol{A}\frac{d\boldsymbol{\psi}(x)}{dx}+\boldsymbol{\chi}(x)=\boldsymbol{q}^{-}(x)$$
$$\frac{1}{2}\boldsymbol{A}\frac{d\boldsymbol{\chi}(x)}{dx}+\boldsymbol{\psi}(x)=\omega\boldsymbol{\delta}\boldsymbol{\psi}(x)+\boldsymbol{q}^{+}(x) \tag{12a,b}$$

On combining Eqs(12) by differentiating Eq(12a)

$$\frac{1}{2}\boldsymbol{A}\frac{d^2\boldsymbol{\psi}(x)}{dx^2}+\frac{d\boldsymbol{\chi}(x)}{dx}=\frac{d\boldsymbol{q}^{-}(x)}{dx}. \tag{13}$$

Then, multiplying by $\boldsymbol{A}/2$

$$\frac{1}{4}\boldsymbol{A}^2\frac{d^2\boldsymbol{\psi}(x)}{dx^2}+\frac{1}{2}\boldsymbol{A}\frac{d\boldsymbol{\chi}(x)}{dx}=\frac{1}{2}\boldsymbol{A}\frac{d\boldsymbol{q}^{-}(x)}{dx} \tag{14a}$$

and introducing Eq(12b)

$$\frac{1}{4}\boldsymbol{A}^2\frac{d^2\boldsymbol{\psi}(x)}{dx^2}-[\boldsymbol{I}-\omega\boldsymbol{\delta}]\boldsymbol{\psi}(x)=\frac{1}{2}\boldsymbol{A}\frac{d\boldsymbol{q}^{-}(x)}{dx}-\boldsymbol{q}^{+}(x) \tag{14b}$$

and simplifying

$$\frac{d^2\boldsymbol{\psi}(x)}{dx^2}-\boldsymbol{\Gamma}^2\boldsymbol{\psi}(x)=2\boldsymbol{A}^{-1}\frac{d\boldsymbol{q}^{-}(x)}{dx}-4\boldsymbol{A}^{-2}\boldsymbol{q}^{+}(x), \tag{14c}$$

where

$$\boldsymbol{\Gamma}^2\equiv4\boldsymbol{A}^{-2}\left[\boldsymbol{I}-\omega\boldsymbol{\delta}\right]. \tag{14d}$$

Once $\boldsymbol{\psi}(x)$ is found from Eq(14b), then from Eq(12a)

$$\boldsymbol{\chi}(x)=-\frac{1}{2}\boldsymbol{A}\frac{d\boldsymbol{\psi}(x)}{dx}+\boldsymbol{q}^{-}(x). \tag{15}$$

The two moments of boundary conditions for the solution of Eq(14b) are found from the incoming flux moments at the boundaries [see Eq(1b)]

$$\phi_l^{+}(0)=\int_0^1 d\mu P_l(2\mu-1)f(\mu)$$
$$\phi_l^{-}(a)=0. \tag{16a,b}$$

The exiting moments will be given by the DPN solution designed to accommodate the half-range condition; therefore, the BCs for $\boldsymbol{\psi}(x)$,

$$\boldsymbol{\psi}(0) \equiv \tilde{\boldsymbol{\phi}}^{+}(0) + \tilde{\boldsymbol{\phi}}^{-}(0)$$
$$\boldsymbol{\psi}(a) \equiv \tilde{\boldsymbol{\phi}}^{+}(a), \qquad (17a,b)$$

are not fully determined. This is where the response matrix enters the analysis.

It should be stated that the approach taken thus far is not unique. Instead of a second order ODE for $\boldsymbol{\psi}(x)$, a second order ODE can equally be found for $\chi(x)$. However, there are additional technical issues with the second approach, and therefore will not be further considered.

**3. Solution to the parity equations**

Our task is to solve the following second order inhomogeneous ODE:

$$\frac{d^2\boldsymbol{\psi}(x)}{dx^2} - \boldsymbol{\Gamma}^2\boldsymbol{\psi}(x) = \boldsymbol{\mathcal{Q}}(x), \qquad (18a)$$

with inhomogeneous term

$$\boldsymbol{\mathcal{Q}}(x) \equiv 2\boldsymbol{A}^{-1}\frac{d\boldsymbol{q}^{-}(x)}{dx} - 4\boldsymbol{A}^{-2}\boldsymbol{q}^{+}(x). \qquad (18b)$$

The solution naturally decomposes into additive solutions to the homogeneous and the particular parts.

$$\boldsymbol{\psi}(x) = \boldsymbol{\psi}_h(x) + \boldsymbol{\psi}_p(x). \qquad (19)$$

We begin with the homogeneous parity equation for $\boldsymbol{\psi}_h(x)$,

$$\frac{d^2\boldsymbol{\psi}_h(x)}{dx^2} - \boldsymbol{\Gamma}^2\boldsymbol{\psi}_h(x) = \mathbf{0}, \qquad (20a)$$

and apply diagonalization to $\boldsymbol{\Gamma}$

$$\boldsymbol{\Gamma} = \boldsymbol{T}diag\{\lambda_k; k = 1,...,N\}\boldsymbol{T}^{-1}, \qquad (20b$$

where $\boldsymbol{T}$ is a matrix of size $N$ whose columns are eigenvectors $\boldsymbol{T}_k$ of $\boldsymbol{\Gamma}$ corresponding to the eigenvalues $\lambda_k$. With diagonalization, Eq(20a) becomes

$$\frac{d^2\boldsymbol{y}(x)}{dx^2} - diag\{\lambda_k^2\}\boldsymbol{y}(x) = \mathbf{0}, \qquad (21a)$$

with

$$\boldsymbol{y}(x) \equiv \boldsymbol{T}^{-1}\boldsymbol{\psi}(x). \qquad (21b)$$

Eqs(21) are scalar ODEs along the diagonal with the convenient solution [5]

$$\boldsymbol{y}(x) \equiv diag\left\{\frac{\sinh(\lambda_k x)}{\sinh(\lambda_k a)}\right\}\boldsymbol{y}(a) + diag\left\{\frac{\sinh(\lambda_k (a-x))}{\sinh(\lambda_k a)}\right\}\boldsymbol{y}(0) \tag{22}$$

chosen to directly incorporate the (unknown) boundary conditions. By re-inserting $\boldsymbol{y}$ from Eq(21b), the solution to Eq(20a) is

$$\boldsymbol{\psi}_h(x) = \boldsymbol{H}(x)\boldsymbol{\psi}_h(a) + \boldsymbol{H}(a-x)\boldsymbol{\psi}_h(0), \tag{23a}$$

where the matrix function $\boldsymbol{H}(x)$ forms the two solutions to the homogeneous part $\boldsymbol{H}(x)$, $\boldsymbol{H}(a-x)$

$$\boldsymbol{H}(x) \equiv \boldsymbol{T} diag\left\{\frac{\sinh(\lambda_k x)}{\sinh(\lambda_k a)}\right\}\boldsymbol{T}^{-1} \tag{23b}$$

$$\boldsymbol{H}(a-x) \equiv \boldsymbol{T} diag\left\{\frac{\sinh(\lambda_k (a-x))}{\sinh(\lambda_k a)}\right\}\boldsymbol{T}^{-1}. \tag{23c}$$

From Eq(19)

$$\begin{aligned}\boldsymbol{\psi}_h(0) &= \boldsymbol{\psi}(0) - \boldsymbol{\psi}_p(0)\\ \boldsymbol{\psi}_h(a) &= \boldsymbol{\psi}(a) - \boldsymbol{\psi}_p(a),\end{aligned} \tag{24a}$$

we find for Eq(23a)

$$\boldsymbol{\psi}(x) - \boldsymbol{\psi}_p(x) = \boldsymbol{H}(x)\left[\boldsymbol{\psi}(a) - \boldsymbol{\psi}_p(a)\right] + \boldsymbol{H}(a-x)\left[\boldsymbol{\psi}(0) - \boldsymbol{\psi}_p(0)\right] \tag{24b}$$

or

$$\boldsymbol{\psi}(x) = \boldsymbol{H}(x)\boldsymbol{\psi}(a) + \boldsymbol{H}(a-x)\boldsymbol{\psi}(0) + \boldsymbol{\psi}_p(x) - \boldsymbol{\psi}_p(a) - \boldsymbol{\psi}_p(0). \tag{24c}$$

Note that the solution contains the unknown boundary conditions.

With the chosen two solutions of Eqs(23b,c) to the homogeneous part and from variation of parameters, the particular solution is [5]

$$\boldsymbol{\psi}_p(x) = \boldsymbol{W}^{-1}\left[\boldsymbol{H}(x)\int_x^a dx'\boldsymbol{H}(a-x')\boldsymbol{Q}(x') + \boldsymbol{H}(a-x)\int_0^x dx'\boldsymbol{H}(x')\boldsymbol{Q}(x')\right] \tag{25a}$$

with Wronskian

$$\boldsymbol{W}^{-1} \equiv -\boldsymbol{T}\left[diag\left\{\frac{sinh(\lambda_k a)}{\lambda_k}\right\}\right]\boldsymbol{T}^{-1}. \tag{25b}$$

Note that the particular solution has been constructed such that

$$\boldsymbol{\psi}_p(0)=\boldsymbol{\psi}_p(a)=0. \tag{26}$$

From Eq(24c)

$$\boldsymbol{\psi}(x)=\boldsymbol{H}(x)\boldsymbol{\psi}(a)+\boldsymbol{H}(a-x)\boldsymbol{\psi}(0)+\boldsymbol{\psi}_p(x); \tag{27a}$$

and from Eq(15)

$$\boldsymbol{\chi}(x)=-\frac{1}{2}\boldsymbol{A}\left[\boldsymbol{H}'(x)\boldsymbol{\psi}(a)+\boldsymbol{H}'(a-x)\boldsymbol{\psi}(0)\right]+\boldsymbol{U}(x), \tag{27b}$$

where

$$\begin{gathered}\boldsymbol{H}'(x)\equiv\boldsymbol{T}diag\left\{\lambda_k\frac{\cosh(\lambda_k x)}{\sinh(\lambda_k a)}\right\}\boldsymbol{T}^{-1}\\ \boldsymbol{H}'(a-x)\equiv-\boldsymbol{T}diag\left\{\lambda_k\frac{\cosh(\lambda_k(a-x))}{\sinh(\lambda_k a)}\right\}\boldsymbol{T}^{-1}\\ \boldsymbol{U}(x)\equiv\boldsymbol{q}^-(x)-\frac{1}{2}\boldsymbol{A}\boldsymbol{\psi}_p'(x).\end{gathered} \tag{27c,d,e}$$

The flux moments in Eqs(2a,b) will be constructed from Eqs(27); but since the parity vectors $\boldsymbol{\psi}(0),\boldsymbol{\psi}(a)$

$$\begin{gathered}\boldsymbol{\psi}(0)=\tilde{\boldsymbol{\phi}}^+(0)+\tilde{\boldsymbol{\phi}}^-(0)\\ \boldsymbol{\psi}(a)=\tilde{\boldsymbol{\phi}}^+(a)+\tilde{\boldsymbol{\phi}}^-(a)\end{gathered} \tag{28}$$

contain the unknown exiting flux vectors $\tilde{\boldsymbol{\phi}}^-(0),\tilde{\boldsymbol{\phi}}^+(a)$, these must be determined first.

## 4. Response matrix

To begin, we define $\hat{\boldsymbol{A}},\hat{\boldsymbol{B}}$

$$\begin{gathered}\boldsymbol{H}'(x)\big|_0\equiv\boldsymbol{T}diag\left\{\frac{\lambda_k}{\sinh(\lambda_k a)}\right\}\boldsymbol{T}^{-1}\equiv\hat{\boldsymbol{B}}\\ \boldsymbol{H}'(x)\big|_a\equiv\boldsymbol{T}diag\left\{\lambda_k\frac{\cosh(\lambda_k a)}{\sinh(\lambda_k a)}\right\}\boldsymbol{T}^{-1}\equiv-\hat{\boldsymbol{A}};\end{gathered} \tag{29a,b}$$

and note the following:

$$\boldsymbol{H}'(a-x)\Big|_0 \equiv -\boldsymbol{T}diag\left\{\lambda_k \frac{\cosh(\lambda_k(a))}{\sinh(\lambda_k a)}\right\}\boldsymbol{T}^{-1} = \hat{\boldsymbol{A}}$$
$$\boldsymbol{H}'(a-x)\Big|_a \equiv -\boldsymbol{T}diag\left\{\frac{\lambda_k}{\sinh(\lambda_k a)}\right\}\boldsymbol{T}^{-1} = -\hat{\boldsymbol{B}}.$$
(29c,d)

To recover the outgoing positive and negative moments at the boundaries, one introduces $x = 0$ and $a$ into Eq(27b) to give

$$\boldsymbol{\chi}(0) = -\frac{1}{2}\boldsymbol{A}\left[\boldsymbol{H}'(x)\Big|_0 \boldsymbol{\psi}(a) + \boldsymbol{H}'(a-x)\Big|_0 \boldsymbol{\psi}(0)\right] + \boldsymbol{U}(0)$$
$$\boldsymbol{\chi}(a) = -\frac{1}{2}\boldsymbol{A}\left[\boldsymbol{H}'(x)\Big|_a \boldsymbol{\psi}(a) + \boldsymbol{H}'(a-x)\Big|_a \boldsymbol{\psi}(0)\right] + \boldsymbol{U}(a)$$
(30a,b)

or from Eqs(28a,b)

$$\boldsymbol{\chi}(0) = -\frac{1}{2}\boldsymbol{A}\left[\hat{\boldsymbol{B}}\boldsymbol{\psi}(a) + \hat{\boldsymbol{A}}\boldsymbol{\psi}(0)\right] + \boldsymbol{U}(0)$$
$$\boldsymbol{\chi}(a) = \frac{1}{2}\boldsymbol{A}\left[\hat{\boldsymbol{A}}\boldsymbol{\psi}(a) + \hat{\boldsymbol{B}}\boldsymbol{\psi}(0)\right] + \boldsymbol{U}(a)$$
(31a,b)

Substituting Eqs(11a,b) at $x = 0$ and $a$ respectively yields

$$\tilde{\boldsymbol{\phi}}^+(0) - \tilde{\boldsymbol{\phi}}^-(0) = -\frac{1}{2}\boldsymbol{A}\left\{\hat{\boldsymbol{B}}\left[\tilde{\boldsymbol{\phi}}^+(a) + \tilde{\boldsymbol{\phi}}^-(a)\right] + \hat{\boldsymbol{A}}\left[\tilde{\boldsymbol{\phi}}^+(0) + \tilde{\boldsymbol{\phi}}^-(0)\right] + \boldsymbol{U}(0)\right\}$$
$$\tilde{\boldsymbol{\phi}}^+(a) - \tilde{\boldsymbol{\phi}}^-(a) = \frac{1}{2}\boldsymbol{A}\left\{\hat{\boldsymbol{A}}\left[\tilde{\boldsymbol{\phi}}^+(a) + \tilde{\boldsymbol{\phi}}^-(a)\right] + \hat{\boldsymbol{B}}\left[\tilde{\boldsymbol{\phi}}^+(0) + \tilde{\boldsymbol{\phi}}^-(0)\right] + \boldsymbol{U}(a)\right\}$$
(32a,b)

and if

$$\boldsymbol{\beta} \equiv \frac{\boldsymbol{A}\hat{\boldsymbol{B}}}{2} \qquad \boldsymbol{\gamma} \equiv \frac{\boldsymbol{A}\hat{\boldsymbol{A}}}{2},$$
(32c,d)

then Eqs(32a,b) become

$$\tilde{\boldsymbol{\phi}}^+(0) - \tilde{\boldsymbol{\phi}}^-(0) = -\boldsymbol{\beta}\left[\tilde{\boldsymbol{\phi}}^+(a) + \tilde{\boldsymbol{\phi}}^-(a)\right] - \boldsymbol{\gamma}\left[\tilde{\boldsymbol{\phi}}^+(0) + \tilde{\boldsymbol{\phi}}^-(0)\right] + \boldsymbol{U}(0)$$
$$\tilde{\boldsymbol{\phi}}^+(a) - \tilde{\boldsymbol{\phi}}^-(a) = \boldsymbol{\gamma}\left[\tilde{\boldsymbol{\phi}}^+(a) + \tilde{\boldsymbol{\phi}}^-(a)\right] + \boldsymbol{\beta}\left[\tilde{\boldsymbol{\phi}}^+(0) + \tilde{\boldsymbol{\phi}}^-(0)\right] + \boldsymbol{U}(a);$$
(32e,f)

and re-arranging

$$\boldsymbol{\beta}\tilde{\boldsymbol{\phi}}^+(a) - [\boldsymbol{I} - \boldsymbol{\gamma}]\tilde{\boldsymbol{\phi}}^-(0) = -\boldsymbol{\beta}\tilde{\boldsymbol{\phi}}^-(a) - [I + \gamma]\tilde{\boldsymbol{\phi}}^+(0) + \boldsymbol{U}(0)$$
$$[\boldsymbol{I} - \boldsymbol{\gamma}]\tilde{\boldsymbol{\phi}}^+(a) - \boldsymbol{\beta}\tilde{\boldsymbol{\phi}}^-(0) = [I + \gamma]\tilde{\boldsymbol{\phi}}^-(a) + \boldsymbol{\beta}\tilde{\boldsymbol{\phi}}^+(0) + \boldsymbol{U}(a)$$
(32g,h)

into matrix form

$$\begin{bmatrix} \boldsymbol{\beta} & -\boldsymbol{x}^- \\ \boldsymbol{x}^- & -\boldsymbol{\beta} \end{bmatrix} \begin{bmatrix} \tilde{\boldsymbol{\phi}}^+(a) \\ \tilde{\boldsymbol{\phi}}^-(0) \end{bmatrix} = \begin{bmatrix} -\boldsymbol{\beta} & -\boldsymbol{x}^+ \\ \boldsymbol{x}^+ & \boldsymbol{\beta} \end{bmatrix} \begin{bmatrix} \tilde{\boldsymbol{\phi}}^-(a) \\ \tilde{\boldsymbol{\phi}}^+(0) \end{bmatrix} + \begin{bmatrix} \boldsymbol{U}(0) \\ \boldsymbol{U}(a) \end{bmatrix}, \tag{32i,j}$$

where

$$x^{\pm} \equiv [\boldsymbol{I} \pm \boldsymbol{\gamma}]. \tag{32k}$$

When we multiply both sides by the skew symmetric block identity matrix

$$\begin{bmatrix} -\boldsymbol{I} & 0 \\ 0 & \boldsymbol{I} \end{bmatrix} \begin{bmatrix} \boldsymbol{\beta} & -\boldsymbol{x}^- \\ \boldsymbol{x}^- & -\boldsymbol{\beta} \end{bmatrix} \begin{bmatrix} \tilde{\boldsymbol{\phi}}^+(a) \\ \tilde{\boldsymbol{\phi}}^-(0) \end{bmatrix} = \begin{bmatrix} -\boldsymbol{I} & 0 \\ 0 & \boldsymbol{I} \end{bmatrix} \begin{bmatrix} -\boldsymbol{\beta} & -\boldsymbol{x}^+ \\ \boldsymbol{x}^+ & \boldsymbol{\beta} \end{bmatrix} \begin{bmatrix} \tilde{\boldsymbol{\phi}}^-(a) \\ \tilde{\boldsymbol{\phi}}^+(0) \end{bmatrix} + \begin{bmatrix} -\boldsymbol{I} & 0 \\ 0 & \boldsymbol{I} \end{bmatrix} \begin{bmatrix} \boldsymbol{U}(0) \\ \boldsymbol{U}(a) \end{bmatrix}, \tag{33a}$$

we find

$$\begin{bmatrix} -\boldsymbol{\beta} & \boldsymbol{x}^- \\ \boldsymbol{x}^- & -\boldsymbol{\beta} \end{bmatrix} \begin{bmatrix} \tilde{\boldsymbol{\phi}}^+(a) \\ \tilde{\boldsymbol{\phi}}^-(0) \end{bmatrix} = \begin{bmatrix} \boldsymbol{\beta} & \boldsymbol{x}^+ \\ \boldsymbol{x}^+ & \boldsymbol{\beta} \end{bmatrix} \begin{bmatrix} \tilde{\boldsymbol{\phi}}^-(a) \\ \tilde{\boldsymbol{\phi}}^+(0) \end{bmatrix} + \begin{bmatrix} -\boldsymbol{U}(0) \\ \boldsymbol{U}(a) \end{bmatrix}. \tag{33b}$$

Finally, multiplying by the inverse of the leading matrix gives the exiting flux

$$\begin{bmatrix} \tilde{\boldsymbol{\phi}}^+(a) \\ \tilde{\boldsymbol{\phi}}^-(0) \end{bmatrix} = \boldsymbol{R} \begin{bmatrix} \tilde{\boldsymbol{\phi}}^-(a) \\ \tilde{\boldsymbol{\phi}}^+(0) \end{bmatrix} + \begin{bmatrix} -\boldsymbol{\beta} & \boldsymbol{x}^- \\ \boldsymbol{x}^- & -\boldsymbol{\beta} \end{bmatrix}^{-1} \begin{bmatrix} -\boldsymbol{U}(0) \\ \boldsymbol{U}(a) \end{bmatrix}, \tag{34a}$$

where the response matrix is

$$\boldsymbol{R} \equiv \begin{bmatrix} -\boldsymbol{\beta} & \boldsymbol{x}^- \\ \boldsymbol{x}^- & -\boldsymbol{\beta} \end{bmatrix}^{-1} \begin{bmatrix} \boldsymbol{\beta} & \boldsymbol{x}^+ \\ \boldsymbol{x}^+ & \boldsymbol{\beta} \end{bmatrix}. \tag{34b}$$

To complete the expression

$$\begin{bmatrix} -\boldsymbol{U}(0) \\ \boldsymbol{U}(a) \end{bmatrix} = \begin{bmatrix} -\boldsymbol{q}^-(0) \\ \boldsymbol{q}^-(a) \end{bmatrix} + \frac{1}{2} \begin{bmatrix} \boldsymbol{\psi}_p'(0) \\ -\boldsymbol{\psi}_p'(a) \end{bmatrix}, \tag{35a,b}$$

with

$$\begin{bmatrix} \boldsymbol{\psi}_p'(0) \\ -\boldsymbol{\psi}_p'(a) \end{bmatrix} = \boldsymbol{W}^{-1} \begin{bmatrix} \hat{\boldsymbol{B}} \int_0^a dx' \boldsymbol{H}(a-x') \boldsymbol{Q}(x') \\ -\hat{\boldsymbol{A}} \int_0^a dx' \boldsymbol{H}(x') \boldsymbol{Q}(x') \end{bmatrix}. \tag{35c,d}$$

On the left, we have the moments of the outgoing flux distribution and on the right the moments of the incoming flux distribution. Hence, for a known incoming distribution, the outgoing moments are now known.

**5. Final moments solution**

One then recovers the $N$ spatial moments from Eqs(11c)$^{\pm}$ as

$$\tilde{\boldsymbol{\phi}}^{+}(x)=\frac{1}{2}\left[\boldsymbol{\psi}(x)+\boldsymbol{\chi}(x)\right]$$
$$\tilde{\boldsymbol{\phi}}^{-}(x)=\frac{1}{2}\left[\boldsymbol{\psi}(x)-\boldsymbol{\chi}(x)\right].$$
(36a)$^{\pm}$

and from Eqs(27a,b) to give the general solution

$$\tilde{\boldsymbol{\phi}}^{+}(x)=\frac{1}{2}\left[\begin{array}{l}\left[\boldsymbol{H}(x)-\frac{1}{2}\boldsymbol{AH}'(x)\right]\boldsymbol{\psi}(a)+\left[\boldsymbol{H}(a-x)-\frac{1}{2}\boldsymbol{AH}'(a-x)\right]\boldsymbol{\psi}(0)+\\ \qquad+\boldsymbol{\psi}_{p}(x)+\boldsymbol{U}(x)\end{array}\right]$$
$$\tilde{\boldsymbol{\phi}}^{-}(x)=\frac{1}{2}\left[\begin{array}{l}\left[\boldsymbol{H}(x)+\frac{1}{2}\boldsymbol{AH}'(x)\right]\boldsymbol{\psi}(a)+\left[\boldsymbol{H}(a-x)+\frac{1}{2}\boldsymbol{AH}'(a-x)\right]\boldsymbol{\psi}(0)-\\ \qquad+\boldsymbol{\psi}_{p}(x)-\boldsymbol{U}(x)\end{array}\right].$$
(36b)$^{\pm}$

Or with

$$\boldsymbol{\mathcal{H}}^{\pm}(x)\equiv\boldsymbol{H}(x)\pm\frac{1}{2}\boldsymbol{AH}'(x)$$
(37a)$^{\pm}$

and

$$\boldsymbol{\mathcal{P}}^{\pm}(x)\equiv\boldsymbol{\psi}_{p}(x)\pm\boldsymbol{U}(x),$$
(37b)$^{\pm}$

we have

$$\tilde{\boldsymbol{\phi}}^{+}(x)=\frac{1}{2}\left[\boldsymbol{\mathcal{H}}^{-}(x)\boldsymbol{\psi}(a)+\boldsymbol{\mathcal{H}}^{-}(a-x)\boldsymbol{\psi}(0)+\boldsymbol{\mathcal{P}}^{+}(x)\right]$$
$$\tilde{\boldsymbol{\phi}}^{-}(x)=\frac{1}{2}\left[\boldsymbol{\mathcal{H}}^{+}(x)\boldsymbol{\psi}(a)-\boldsymbol{\mathcal{H}}^{+}(a-x)\boldsymbol{\psi}(0)\right]+\boldsymbol{\mathcal{P}}^{-}(x).$$
(37c)$^{\pm}$

With additional manipulation

$$\begin{bmatrix}\tilde{\boldsymbol{\phi}}^{+}(x)\\ \tilde{\boldsymbol{\phi}}^{-}(x)\end{bmatrix}=\frac{1}{2}\begin{bmatrix}\boldsymbol{\mathcal{H}}^{-}(x) & \boldsymbol{\mathcal{H}}^{-}(a-x)\\ \boldsymbol{\mathcal{H}}^{+}(x) & \boldsymbol{\mathcal{H}}^{+}(a-x)\end{bmatrix}\left\{\begin{bmatrix}\tilde{\boldsymbol{\phi}}^{+}(a)\\ \tilde{\boldsymbol{\phi}}^{-}(0)\end{bmatrix}+\begin{bmatrix}\tilde{\boldsymbol{\phi}}^{-}(a)\\ \tilde{\boldsymbol{\phi}}^{+}(0)\end{bmatrix}\right\}+\frac{1}{2}\left\{\boldsymbol{\psi}_{p}(x)\begin{bmatrix}\boldsymbol{1}\\ \boldsymbol{1}\end{bmatrix}+\boldsymbol{U}(x)\begin{bmatrix}\boldsymbol{1}\\ \boldsymbol{-1}\end{bmatrix}\right\},$$
(38a)$^{\pm}$

or introducing Eq(34a)

$$\begin{bmatrix}\tilde{\boldsymbol{\phi}}^{+}(x)\\ \tilde{\boldsymbol{\phi}}^{-}(x)\end{bmatrix}=\frac{1}{2}\begin{bmatrix}\boldsymbol{\mathcal{H}}^{-}(x) & \boldsymbol{\mathcal{H}}^{-}(a-x)\\ \boldsymbol{\mathcal{H}}^{+}(x) & \boldsymbol{\mathcal{H}}^{+}(a-x)\end{bmatrix}\left\{\left[\boldsymbol{R}+\boldsymbol{I}\right]\begin{bmatrix}\tilde{\boldsymbol{\phi}}^{-}(a)\\ \tilde{\boldsymbol{\phi}}^{+}(0)\end{bmatrix}+\begin{bmatrix}-\boldsymbol{\beta} & \boldsymbol{x}^{-}\\ \boldsymbol{x}^{-} & -\boldsymbol{\beta}\end{bmatrix}^{-1}\begin{bmatrix}-\boldsymbol{U}(0)\\ \boldsymbol{U}(a)\end{bmatrix}\right\}+$$
$$+\frac{1}{2}\left\{\boldsymbol{\psi}_{p}(x)\begin{bmatrix}\boldsymbol{1}\\ \boldsymbol{1}\end{bmatrix}+\boldsymbol{U}(x)\begin{bmatrix}\boldsymbol{1}\\ \boldsymbol{-1}\end{bmatrix}\right\}$$
(38b)$^{\pm}$

with

$$\begin{bmatrix} \tilde{\boldsymbol{\phi}}^{-}(a) \\ \tilde{\boldsymbol{\phi}}^{+}(0) \end{bmatrix} = \begin{bmatrix} \mathbf{0} \\ \int_0^1 d\mu \boldsymbol{P}_N(2\mu-1) f(\mu) \end{bmatrix}. \quad (38c)^{\pm}$$

and

$$\boldsymbol{P}_N(2\mu-1) \equiv \left[P_0(2\mu-1) \quad P_1(2\mu-1) \quad \dots \quad P_N(2\mu-1)\right]^T$$
$$\boldsymbol{1} \equiv \{1\ 1\ \dots .1\}^T. \quad (38d,e)$$

and $\boldsymbol{U}(x)$ is from Eq(27e).

**6. The DPN Approximation**

The DPN solution comes from the convergence of the series solution of Eqs(2)

$$\phi(x,\mu) = \lim_{N\to\infty} \phi(x,\mu;N), \quad (39a)$$

where the DPN approximation is the partial sum

$$\phi(x,\mu;N) = \begin{cases} \sum_{l=0}^{N} (2l+1)\phi_l^{+}(x) P_l(2\mu-1), \ \mu \ge 0 \\ \sum_{l=0}^{N} (2l+1)\phi_l^{-}(x) P_l(2\mu+1), \ \mu \le 0. \end{cases} \quad (39b,c)$$

To incorporate the moments vector of Eqs(38b), we let $\mu$ negative be $-|\mu|$ and trivially multiply by $(+1)^l$ in the partial sum for $\mu$ positive to give for $\mu \ge 0$

$$\phi(x,\pm\mu;N) = \sum_{l=0}^{N} (2l+1)\tilde{\phi}_l^{\pm}(x) P_l(2|\mu|-1), \quad (40a)$$

and if

$$L_N \equiv diag\{2l+1,\ l=1,\dots,N\}, \quad (40b)$$

Eq(40a) becomes the $N$-term inner product

$$\phi(x,\pm\mu;N) = \boldsymbol{P}_N^T(2|\mu|-1)\boldsymbol{L}_N \tilde{\boldsymbol{\phi}}^{\pm}(x) \quad (40c)$$

to be evaluated in the next section.

**7. Numerical implementation and demonstration**

**7a. Numerical implementation for an isotropic source**

Numerical implementation of the DPN algorithm for an isotropic source at the near surface and none at the far surface is relatively straightforward by following the algorithm presented. In particular, matrix diagonalization is through an HQR procedure originating from the LAPACK routine [6] and is one of the most reliable of numerical methods in use today. Once eigenvalues are known, the matrix functions come from common matrix multiplication of Eqs(23b,c). LU decomposition gives the matrix inversions. Our final consideration is convergence in DPN order $N$ through a sequence of DPN approximations [Eq(40c)] and convergence acceleration. To develop a sequence of partial sums, we increment the number of moments by stride $\Delta N$ (usually $\Delta N = 5$) and monitor convergence of the sequence through sequential convergence and convergence acceleration until the relative error is below a desired relative error or convergence fails. If failure, we increase the number of sequence elements or change expectations. In this way, convergence provides a measure of closure. There are numerous variants of acceleration to choose from. Here, we first apply common sequential convergence similar to a sensitivity study and then Wynn-epsilon (W-e) acceleration. Sequential convergence starts from DPN order $N_0$ and compares the convergence standard, called the engineering estimate of the relative error between iterations $m-1$ and $m$ [i.e., orders $N_0+(m-1)\Delta N$ and $N_0+m\Delta N$], for $m$=1,2,…

$$\varepsilon_{Seq}(x,\mu;m) \equiv \left| \frac{\phi(x,\pm\mu;N_0+m\Delta N)-\phi(x,\pm\mu;N_0+(m-1)\Delta N)}{\phi(x,\pm\mu;N_0+m\Delta N)} \right|, \qquad (41a)$$

to the desired relative error $\varepsilon$

$$\varepsilon_{Seq}(x,\mu;m) < \varepsilon. \qquad (41b)$$

If satisfied, the sequence has converged sequentially at $(x,\mu)$. *W-e* acceleration, in contrast, is non-linear. Similarly, to sequential convergence, the same sequence of DPN approximations

$$s_m = \phi(x,\pm\mu;N_0+m\,\Delta N),\ m=0,1,... \qquad (42a)$$

is the initial sequence list. W-e convergence essentially extrapolates a known sequence to give the next sequence element (and an estimate of the limit) from previous elements. The algorithm is recursive and for $L$+1 iterations, in principle, improved DPN partial sums result for $k$ odd

$$\begin{aligned}
&\varepsilon_{-1}^{(m)} = 0\\
&\varepsilon_{0}^{(m)} = s_m,\ m=0,1,...,L\\
&\varepsilon_{k+1}^{(m)} = \varepsilon_{k-1}^{(m+1)} + \left[\varepsilon_{k}^{(m+1)} - \varepsilon_{k}^{(m)}\right]^{-1};\ m=0,1,...,L-k-1;\ k=0,1,...,L-1.
\end{aligned} \qquad (42b)$$

The algorithm is conveniently written as the following tableau:

$$\begin{array}{cccccc}
\varepsilon_0^{(0)} & \varepsilon_1^{(0)} & \varepsilon_2^{(0)} & \dots & \varepsilon_{L-1}^{(0)} & \varepsilon_L^{(0)}\\
\varepsilon_0^{(1)} & \varepsilon_1^{(1)} & \varepsilon_2^{(1)} & \dots & \varepsilon_{L-1}^{(1)} & \\
\varepsilon_0^{(2)} & \dots & & \dots & & \\
\dots & & \varepsilon_2^{(L-2)} & & & \\
 & \varepsilon_1^{(L-1)} & & & & \\
\varepsilon_0^{(L)} & & & & &
\end{array} \qquad (43)$$

Every element of the odd columns should give an improved estimate of the original DPN partial sum in column one. The last term in each column, indicated by the arrow, assumed most precise, gives the following relative error standard after $L$ iterations:

$$e_{We}(x,\mu;L) \equiv \left| \frac{\varepsilon_L^{(0)}(x,\mu) - \varepsilon_{L-2}^{(2)}(x,\mu)}{\varepsilon_L^{(0)}(x,\mu)} \right|, \quad (44a)$$

considered converged when

$$\varepsilon_{We}(x,\mu;L) < \varepsilon;\ L = 1,2,.... \quad (44b)$$

Overall convergence therefore is a competition between the two modes of partial sum convergence, where convergence occurs for the least relative error

$$min\left[\varepsilon_{Seq}(x,\mu;m), \varepsilon_{We}(x,\mu;L)\right] < \varepsilon. \quad (45)$$

As a first demonstration, we consider an isotropic source, $f(\mu) = 1$ entering the near surface and no incoming flux entering the far surface to give surface flux moments vectors

$$\begin{bmatrix} \tilde{\boldsymbol{\phi}}^-(a) \\ \tilde{\boldsymbol{\phi}}^+(0) \end{bmatrix} = \begin{bmatrix} \mathbf{0} \\ \boldsymbol{1}_0 \end{bmatrix}. \quad (46)$$

From Eqs(34a,b), the exiting flux moments vectors are therefore

$$\begin{bmatrix} \tilde{\boldsymbol{\phi}}^+(a) \\ \tilde{\boldsymbol{\phi}}^-(0) \end{bmatrix} = \boldsymbol{R} \begin{bmatrix} \tilde{\boldsymbol{\phi}}^-(a) \\ \tilde{\boldsymbol{\phi}}^+(0) \end{bmatrix} = \begin{bmatrix} -\boldsymbol{\beta} & \boldsymbol{x}^- \\ \boldsymbol{x}^- & -\boldsymbol{\beta} \end{bmatrix}^{-1} \begin{bmatrix} \boldsymbol{\beta} & \boldsymbol{x}^+ \\ \boldsymbol{x}^+ & \boldsymbol{\beta} \end{bmatrix} \begin{bmatrix} \mathbf{0} \\ \boldsymbol{1}_0 \end{bmatrix}, \quad (47a)$$

and the interior flux moment vectors are from Eqs(38b,c)

$$\begin{bmatrix} \tilde{\boldsymbol{\phi}}^+(x) \\ \tilde{\boldsymbol{\phi}}^-(x) \end{bmatrix} = \frac{1}{2} \begin{bmatrix} \boldsymbol{\mathcal{H}}^-(x) & \boldsymbol{\mathcal{H}}^-(a-x) \\ \boldsymbol{\mathcal{H}}^+(x) & \boldsymbol{\mathcal{H}}^+(a-x) \end{bmatrix} [\boldsymbol{R} + \boldsymbol{I}] \begin{bmatrix} \tilde{\boldsymbol{\phi}}^-(a) \\ \tilde{\boldsymbol{\phi}}^+(0) \end{bmatrix}. \quad (47b)$$

The flux vectors enter Eq(40a) re-written as

$$\phi(x,\mu;N) = \sum_{l=0}^{N} a_l^{\pm}(x) \theta_l(\mu) \quad (48a)$$

with

$$\theta_l(\mu) = P_l(2|\mu| - 1)$$
$$a_l^{\pm}(x) \equiv (2l+1) \tilde{\phi}_l^{\pm}(x). \quad (48b)$$

Since $\theta_l(\mu)$ obeys the recurrence

$$\theta_{l+1}(\mu) == (2|\mu|-1)\frac{2l+1}{l}\theta_l(\mu) - \frac{l}{l+1}\theta_{l-1}(\mu), \tag{49a}$$

the Clenshaw sum [7] applies, where

$$\phi(x,\mu;N) = \theta_0(\mu)a_0^{\pm}(x) + \theta_1(\mu)b_1(x,\mu) + \beta_1(\mu)\theta_0(\mu)b_2(x,\mu). \tag{49b}$$

and

$$\begin{aligned}\alpha_l(\mu) &= (2\mu-1)\frac{2l+1}{l} \\ \beta_l(\mu) &= -\frac{l}{l+1}\end{aligned} \tag{49c}$$

$$\begin{aligned}&b_{l+1}(x,\mu) = b_{l+2}(x,\mu) = 0 \\ &b_l(x,\mu) = a_l^{\pm}(x) + \alpha_l(\mu)b_{l+1}(x,\mu) + \beta_{l+1}(\mu)b_{l+2}(x,\mu).\end{aligned} \tag{49d}$$

We now turn to the issue of solution precision. To address precision, we compare the DPN solution to a well-established fully analytical discrete ordinates response matrix benchmark RM/DOM [5]. We assume no absorption ($\omega = 1$) and a slab of 1 *mfp* thickness. Table 1 gives what is believed to be precise angular fluxes to better than one unit in the 8th place using RM/DOM. The DPN angular flux approximation for $N = 100$ agrees to all 7-digits of the angular flux from the RM/DOM of quadrature order 250 and to all but three entries (last digit emboldened and underlined) in the 8th place for $N = 150$ without acceleration. Thus, the method presented indeed does successfully provide an extreme benchmark to nearly 8 places.

The effectiveness of W-e acceleration is shown by Fig. 1, which is the ratio of the relative error with and without acceleration over all directions at the seven spatial coordinates. One observes the W-e relative errors at convergence are generally smaller than the errors of the original sequence for this benchmark solution. This is further confirmation of the high quality-- to one unit in the seventh and nearly one unit in the eighth place-- of the proposed DPN algorithm. 40 of the 72 fluxes converged by W-e showing the significance of the Wynn-epsilon algorithm.

Table 1. Angular Flux for an entering isotropic source.

| μ\x | 0.0 | 0.05 | 0.1 | 0.2 | 0.5 | 0.75 | 1.0 |
|---|---|---|---|---|---|---|---|
| -1.000E+00 | 3.4132876**0**E-01 | 3.20920611E-01 | 3.01041128E-01 | 2.62366118E-01 | 1.53240509E-01 | 7.07430107E-02 | 0.00000000E+00 |
| -8.000E-01 | 3.92084430E-01 | 3.69683964E-01 | 3.47820410E-01 | 3.05071361E-01 | 1.82180798E-01 | 8.6021443**6**E-02 | 0.00000000E+00 |
| -6.000E-01 | 4.58134371E-01 | 4.33685363E-01 | 4.09782681E-01 | 3.62759573E-01 | 2.24012630E-01 | 1.09614879E-01 | 0.00000000E+00 |
| -4.000E-01 | 5.43854301E-01 | 5.17792855E-01 | 4.92356464E-01 | 4.42065792E-01 | 2.88493773E-01 | 1.50567243E-01 | 0.00000000E+00 |
| -2.000E-01 | 6.45967494E-01 | 6.19276078E-01 | 5.93756446E-01 | 5.43978308E-01 | 3.90966211E-01 | 2.35919292E-01 | 0.00000000E+00 |
| 0.000E+00 | 7.58146459E-01 | 7.22978545E-01 | 6.94563136E-01 | 6.42872374E-01 | 5.00000000E-01 | 3.81715377E-01 | 2.41853541E-01 |
| 2.000E-01 | 1.00000000E+00 | 9.42160751E-01 | 8.90352375E-01 | 8.02157479E-01 | 6.09033789E-01 | 4.80750231E-01 | 3.54032506E-01 |
| 4.000E-01 | 1.00000000E+00 | 9.69316928E-01 | 9.38639544E-01 | 8.78613253E-01 | 7.11506227E-01 | 5.83062169E-01 | 4.56145699E-01 |
| 6.000E-01 | 1.00000000E+00 | 9.79131021E-01 | 9.57479617E-01 | 9.12982318E-01 | 7.75987370E-01 | 6.60525697E-01 | 5.41865629E-01 |
| 8.000E-01 | 1.00000000E+00 | 9.84189969E-01 | 9.67479934E-01 | 9.32267600E-01 | 8.17819202E-01 | 7.15927250E-01 | 6.07915570E-01 |
| 1.000E+00 | 1.00000000E+00 | 9.87275159E-01 | 9.73675082E-01 | 9.44578242E-01 | 8.46759491E-01 | 7.56515359E-01 | 6.5867124**0**E-01 |

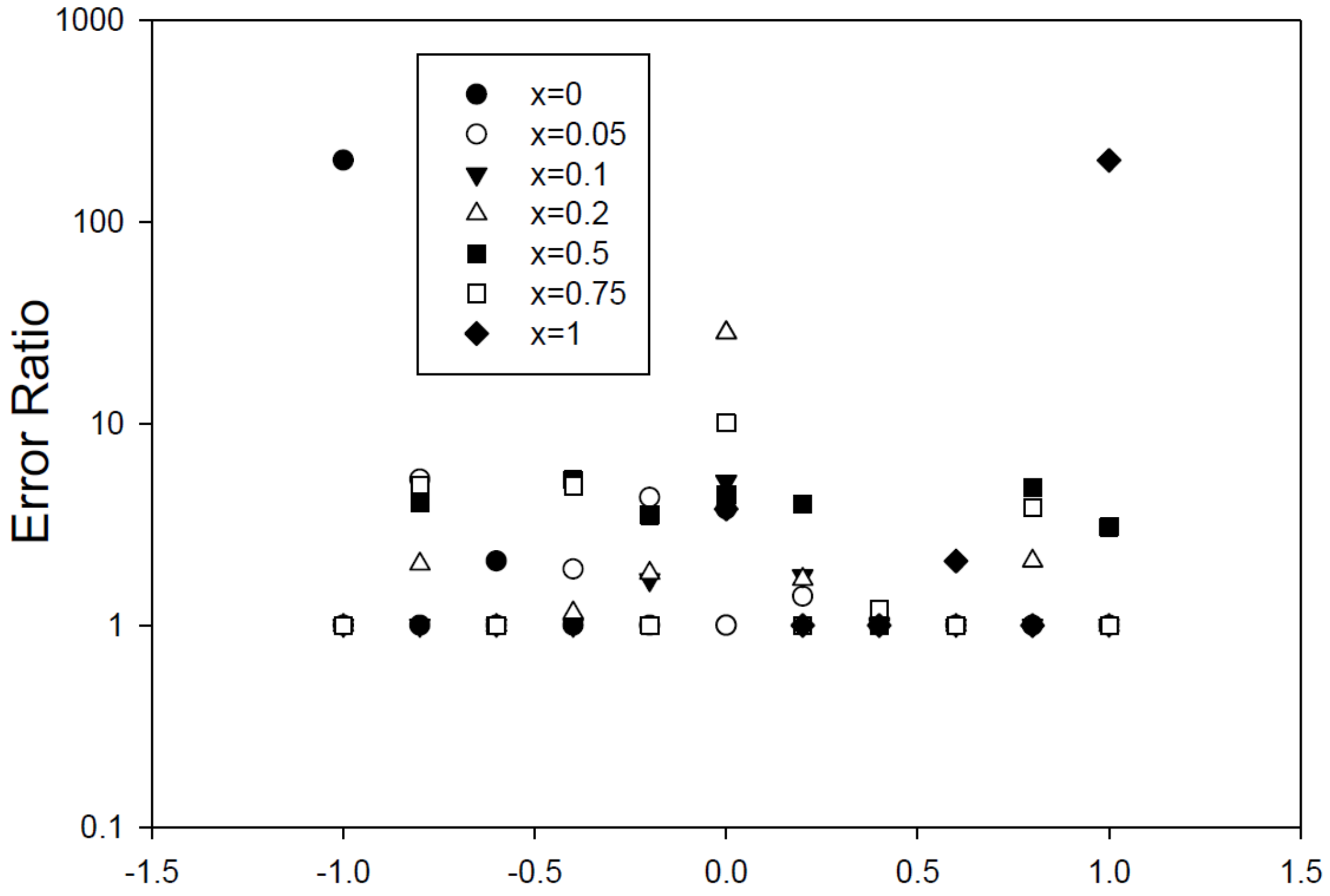

Fig. 1. Ratio of unaccelerated to accelerated relative errors.

**7b. Numerical implementation for a beam source**

The final example is for a beam source entering the near surface and none at the far surface

$$\phi(0,\mu) = f(\mu) = \delta(\mu - \mu_0)$$
$$\phi(a,-\mu) = 0. \tag{50a,b}$$

For this case, it is convenient to separate the uncollided $\phi_0(x,\mu)$ from the collided $\phi_c(x,\mu)$ component

$$\phi(x,\mu) = \phi_0(x,\mu) + \phi_c(x,\mu), \tag{51}$$

where he uncollided (scalar) component should not be confused with the zeroth vector flux moments $\tilde{\phi}_0^{\pm}(x)$, a vector. The uncollided flux satisfies

$$\left[\mu\frac{\partial}{\partial x} + 1\right]\phi_0(x,\mu) = 0 \tag{52a}$$

$$\phi_0(0,\mu) = \delta(\mu - \mu_0)$$
$$\phi_0(a,-\mu) = 0 \tag{52b}$$

to give

$$\phi_0(x,\mu) = \delta(\mu - \mu_0)e^{-x/\mu_0}\Theta(\mu) \tag{52c}$$

where $\Theta(\mu)$ is the Heaviside step function required to maintain the uncollided flux in the positive direction. When Eq(51) is introduced into Eq(1a) with Eq(52c)

$$\left[\mu\frac{\partial}{\partial x}+1\right]\phi_c(x,\mu)=\frac{\omega}{2}\int_{-1}^{1}d\mu'\phi_c(x,\mu')+Q(x)$$
$$Q(x)\equiv\frac{\omega}{2}e^{-x/\mu_0},\tag{53a,b}$$

with boundary conditions

$$\begin{aligned}\phi_c(0,\mu)&=0\\ \phi_c(a,-\mu)&=0.\end{aligned}\tag{53c,d}$$

From Eq(14c) for the collided component derived from Eq(53b)

$$\frac{d^2\boldsymbol{\psi}_c(x)}{dx^2}-\boldsymbol{\Gamma}^2\boldsymbol{\psi}_c(x)=-4\omega e^{-x/\mu_0}\boldsymbol{A}^{-2}\boldsymbol{1}_0\tag{54a}$$

with

$$\begin{aligned}\boldsymbol{\psi}_c(0)&=\tilde{\boldsymbol{\phi}}_c^-(0)\\ \boldsymbol{\psi}_c(a)&=\tilde{\boldsymbol{\phi}}_c^+(a).\end{aligned}\tag{54b,c}$$

Assuming the particular solution

$$\boldsymbol{\psi}_p(x)=e^{-x/\mu_0}\boldsymbol{C},\tag{55a}$$

we find

$$\boldsymbol{C}\equiv-4\mu_0^2\omega\left[I-\mu_0^2\boldsymbol{\Gamma}^2\right]\boldsymbol{A}^{-2}\boldsymbol{1}_0.\tag{55b}$$

From Eq(38b)

$$\begin{bmatrix}\tilde{\boldsymbol{\phi}}_c^+(x)\\ \tilde{\boldsymbol{\phi}}_c^-(x)\end{bmatrix}=\frac{1}{2}\begin{bmatrix}\boldsymbol{\mathcal{H}}^-(x) & \boldsymbol{\mathcal{H}}^-(a-x)\\ \boldsymbol{\mathcal{H}}^+(x) & \boldsymbol{\mathcal{H}}^+(a-x)\end{bmatrix}\begin{bmatrix}\tilde{\boldsymbol{\phi}}_c^+(a)\\ \tilde{\boldsymbol{\phi}}_c^-(0)\end{bmatrix}+\frac{1}{2}\left\{\boldsymbol{\psi}_p(x)\begin{bmatrix}\boldsymbol{1}\\ \boldsymbol{1}\end{bmatrix}+\frac{e^{-x/\mu_0}}{2\mu_0}\boldsymbol{A}\begin{bmatrix}\boldsymbol{1}\\ -\boldsymbol{1}\end{bmatrix}\right\},\tag{56a}$$

where

$$\begin{bmatrix}\tilde{\boldsymbol{\phi}}_c^+(a)\\ \tilde{\boldsymbol{\phi}}_c^-(0)\end{bmatrix}=\begin{bmatrix}-\boldsymbol{\beta} & \boldsymbol{x}^-\\ \boldsymbol{x}^- & -\boldsymbol{\beta}\end{bmatrix}^{-1}\begin{bmatrix}-\boldsymbol{U}(0)\\ \boldsymbol{U}(a)\end{bmatrix},\tag{56b}$$

and

$$\begin{bmatrix} -\boldsymbol{U}(0) \\ \boldsymbol{U}(a) \end{bmatrix} = \frac{1}{2}\begin{bmatrix} \boldsymbol{\psi}'_p(0) \\ -\boldsymbol{\psi}'_p(a) \end{bmatrix} = -\frac{1}{2\mu_0}\begin{bmatrix} \boldsymbol{I} & \boldsymbol{0} \\ \boldsymbol{0} & e^{-a/\mu_0}\boldsymbol{I} \end{bmatrix}\begin{bmatrix} \boldsymbol{C} \\ -\boldsymbol{C} \end{bmatrix} \tag{56c}$$

to be inserted into

$$\phi_c(x, \pm\mu; N) = \boldsymbol{P}_N^T(2|\mu|-1)\boldsymbol{L}_N\tilde{\boldsymbol{\phi}}_c^{\pm}(x) \tag{56d}$$

for the angular flux.

The slab for the beam source is the same as for the isotropic source. Table 2 shows the results from RM/DOM [5], which is a 7-place benchmark. Overlaid is the DPN result where it is observed that DPN gives nearly all 7 places except for two entries in one unit in the last digit. The beam source behaves differently from the isotropic source however, as it is more sensitive to round off error. The DPN results of table 2 required quadruple precision (QP) to achieve nearly 7-places; whereas, the isotropic source required double precision. For the beam source, double precision only gives at best 4-places. A future effort will attempt to determine where the loss of precision occurs; nevertheless, the DPN provides a solid 6 digits and near 7 with QP.

Table 2. Angular Flux for a perpendicularly ($\mu_0$ =1) entering beam source.

| $\mu \backslash x$ | 0 | 0.05 | 0.1 | 0.2 | 0.5 | 0.75 | 1 |
|---|---|---|---|---|---|---|---|
| -1.000E+00 | 5.3877491E-01 | 5.1979897E-01 | 4.9826415E-01 | 4.5015758E-01 | 2.8363970E-01 | 1.3670184E-01 | 0.0000000E+00 |
| -8.000E-01 | 6.1358488E-01 | 5.9454278E-01 | 5.7227580E-01 | 5.2122894E-01 | 3.3675659E-01 | 1.6617467E-01 | 0.0000000E+00 |
| -6.000E-01 | 7.0705901E-01 | 6.8953074E-01 | 6.6778458E-01 | 6.1558120E-01 | 4.1317580E-01 | 2.1164480E-01 | 0.0000000E+00 |
| -4.000E-01 | 8.1805757E-01 | 8.0600276E-01 | 7.8820201E-01 | 7.4066647E-01 | 5.2986404E-01 | 2.9042486E-01 | 0.0000000E+00 |
| -2.000E-01 | 9.1606674E-01 | 9.1832709E-01 | 9.1231716E-01 | 8.8438367E-01 | 7.1013555E-01 | 4.5375071E-01 | 0.0000000E+00 |
| 0.000E+00 | 8.7868708E-01 | 9.2867409E-01 | 9.4863473E-01 | 9.5773321E-01 | 8.6865051E-01 | 7.1731075E-01 | 4.8302802E-01 |
| 2.000E-01 | 0.0000000E+00 | 2.0129990E-01 | 3.6476856E-01 | 5.9744092E-01 | 8.4223675E-01 | 7.9950036E-01 | 6.5012804E-01 |
| 4.000E-01 | 0.0000000E+00 | 1.0687614E-01 | 2.0478323E-01 | 3.7093644E-01 | 6.5966190E-01 | 7.2022080E-01 | 6.6508565E-01 |
| 6.000E-01 | 0.0000000E+00 | 7.2711601E-02 | 1.4205915E-01 | 2.6699419E-01 | 5.2395290E-01 | 6.1548983E-01 | 6.1208033E-01 |
| 8.000E-01 | 0.0000000E+00 | 5.5092891E-02 | 1.0870690E-01 | 2.0824783E-01 | 4.3113588E-01 | 5.2849359E-01 | 5.4968216E-01 |
| 1.000E+00 | 0.0000000E+00 | 4.4345690E-02 | 8.8026420E-02 | 1.7060822E-01 | 3.6524695E-01 | 4.6023695E-01 | 4.9305877E-01 |

**Conclusion**

By expressing the solution to the 1D monoenergetic neutron transport equation in plane geometry in an infinite series of half-range Legendre polynomials, a first order coupled set of ODEs for half-range moments in the positive and negative directions follows on truncation. By adding and subtracting, the ODEs transform into a second order form for the even and odd parity moments. At this point, one can choose to follow several solution scheme to solve for the parity moments. Here, we choose to establish the even parity solution to include the unknown boundary conditions. By manipulation of the matrix equations, one finds the relationship between the incoming and the exiting fluxes through the response matrix. With the exiting fluxes known, the fluxes interior to the slab also become known. We form the solution in terms of the diagonalization of the matrix associated with the second order ODE, or, in other words, expressible as eigenvectors and eigenvalues. Convergence of the infinite series is through sequential convergence of the partial sums and Wynn-epsilon convergence acceleration. It was shown, via two examples, that benchmarks of 6 and 7 digits can be constructed. It should be noted that the precision quoted depends on the spatial positions and directions sampled but is thought to be representative of a wide variety of samples.